\documentclass[11pt]{article}

\usepackage[final]{acl}

\usepackage{times}
\usepackage{latexsym}

\usepackage[T1]{fontenc}

\usepackage[utf8]{inputenc}

\usepackage{microtype}

\usepackage{inconsolata}

\usepackage{graphicx}

\usepackage{multirow}
\usepackage{booktabs}
\usepackage{tcolorbox}
\usepackage[table]{xcolor}
\tcbuselibrary{breakable,listings}
\usepackage[dvipsnames]{xcolor}
\usepackage{pifont}

\newtcblisting{llmprompttemplate}[1]{
    breakable,
    colback=gray!5,        
    colframe=black!75,     
    fonttitle=\bfseries,   
    coltitle=white,        
    title=#1,              
    boxrule=0.5pt,         
    left=6pt,              
    right=6pt,
    top=6pt,
    bottom=6pt,
    listing only,          
    listing options={
        basicstyle=\ttfamily\small, 
        breaklines=true,            
        columns=fullflexible
    }
}

\newcommand{\scone}{discussion}
\newcommand{\Scone}{Discussion}

\newcommand{\Sctwo}{Turn-based Game}

\newcommand{\textlm}[1]{\textcolor{gray!80}{#1}}

\title{MP-Bench: Evaluating Voice Agents as a Multiparty Conversation Participant}

\author{
 \textbf{Yi-Jen Shih\textsuperscript{1}\thanks{Equal contribution.}},
 \textbf{Shih-Yun Shan Kuan\textsuperscript{2}\footnotemark[1]},
 \textbf{Guan-Ting Lin\textsuperscript{2}\footnotemark[1]},
 \textbf{Kai-Wei Chang\textsuperscript{3}},
\\
 \textbf{Siddhant Arora\textsuperscript{4}},
 \textbf{Shu-wen Yang\textsuperscript{2}},
 \textbf{Abdelrahman Mohamed\textsuperscript{5}},
\\
\textbf{Shinji Watanabe \textsuperscript{4}},
 \textbf{Hung-yi Lee\textsuperscript{2}},
 \textbf{David Harwath\textsuperscript{1}}
\\
\\
 \textsuperscript{1}The University of Texas at Austin,
 \textsuperscript{2}National Taiwan University,
 \\
 \textsuperscript{3}Massachusetts Institute of Technology,
 \textsuperscript{4}Carnegie Mellon University,
 \textsuperscript{5}Meta AI
\\
 \small{
   \textbf{Correspondence:} \href{mailto:yjshih@utexas.edu}{yjshih@utexas.edu}, \href{mailto:r14942094@ntu.edu.tw}{r14942094@ntu.edu.tw}
 }
}

\begin{document}
\maketitle
\begin{abstract}
Conversational voice agents have advanced significantly, offering increasingly natural human-machine interactions through both cascaded and end-to-end architectures. However, while recent benchmarks extensively evaluate dyadic interactions and passive audio comprehension, they largely overlook a prevalent real-world scenario: multiparty conversations. Evaluating agents in these settings is fundamentally more challenging than in dyadic interactions due to the dramatically greater conversational complexity. For voice agents to integrate seamlessly into human group dynamics, they must not only generate contextually appropriate responses but also demonstrate a nuanced understanding of open turn-taking. To address this gap, we introduce Multiparty Bench (MP-Bench), the first benchmark specifically designed to objectively evaluate conversational speech systems as active participants within multiparty contexts. MP-Bench assesses agent behavior along two primary dimensions: turn-taking awareness and response appropriateness. Additionally, we incorporate comprehension-based question-answering tasks as a complementary evaluation. By benchmarking 12 voice agents, we find that realtime voice agents stay at or below $33$\% on multiparty comprehension and remain near chance on multiparty turn-taking, exposing an open challenge for realtime voice agents under multiparty scenario.
\end{abstract}

\section{Introduction}
Conversational voice agents have gained considerable attention in both academia~\cite{wang_2024_freezeomni,wang26fullduplexinteractionspokendialogue} and industry~\cite{google_2025_geminiliveapi,openai_2025_gptrealtime} in recent years, as speech serves as the most natural and intuitive modality for human-machine interaction. 
Furthermore, compared to text, the speech modality conveys rich paralinguistic information, such as speaker identity, emotion, and prosody. 
These systems are designed to process human speech and generate responses in the same audio modality.

\begin{figure}[t]
  \includegraphics[width=\linewidth,trim={0.7cm 0.5cm 0.7cm 0.4cm},clip]{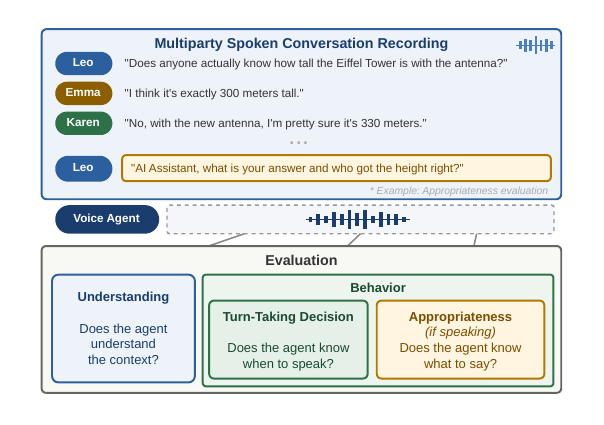}
  \caption {Overview of the evaluation aspects in MP-Bench. Given a prepared multiparty spoken conversation, we evaluate the voice agent's understanding and behavioral abilities.}
  \label{fig:page1fig}
\end{figure}
Based on their interactive behavior, conversational agents can be broadly classified as either non-realtime or realtime models.
Non-realtime agents~\cite{xu_2025_qwen3omni} can comprehend spoken utterances and synthesize audio responses; however, they exhibit considerable latency, precluding seamless, human-like conversations. Real-time models, conversely, are explicitly optimized for interactive dialogue. 
To implement these agents, there are two primary architectural approaches: cascaded and end-to-end. 
The cascaded approach relies on a pipeline comprising a front-end Automatic Speech Recognition (ASR) module, a text-based LLM, and a Text-to-Speech (TTS) synthesizer. The end-to-end approach, on the other hand, employs a single unified model that directly processes speech input and generates speech output, bypassing the need to convert speech to intermediate text representations.

To evaluate the performance of these systems, recent studies have introduced various benchmarks~\cite{arora2025_slm_survey}.
Though multiple dimensions of conversational behavior have been evaluated, existing benchmarks largely overlook multiparty conversations, which are a prevalent scenario in daily human communication.
Ultimately, we envision integrating voice agents into human group interactions, enabling them to participate and contribute to discussions as naturally as humans.

To this end, we introduce Multiparty Bench (MP-Bench) to evaluate voice agents' turn-taking and response behavior in controlled multiparty scenarios spanning explicit, implicit, and negative turn-taking conditions, as a first step toward full participation in natural multiparty conversation~(Fig~\ref{fig:page1fig}).
Our benchmark assesses agent performance along two primary dimensions: turn-taking awareness (knowing when to speak or remain silent) and contextual appropriateness. 
Unlike dyadic interactions where the end of an utterance typically serves as an implicit cue to respond,
multiparty turn-taking introduces more complexity. 
To evaluate these dynamics objectively, we utilize synthetically generated recordings, enabling precise control over the conversational state.

Specifically, we design two scenarios: \textit{\Scone}, where the agent mediates a multiparty dispute, and \textit{\Sctwo}, which requires the agent to adhere to game rules and speak only on its designated turn. 
These scenarios span three distinct turn-taking conditions: \textbf{Explicit} (the agent is addressed by name), \textbf{Implicit} (contextual cues imply it is the agent's turn without explicit naming), and \textbf{Negative} (the agent must correctly remain silent while not their turn). 
Furthermore, we incorporate comprehension tasks to test accurate speaker attribution. 
While our evaluation focuses primarily on realtime voice agents, we also benchmark non-realtime systems and text LLMs for comparison.

Our evaluation reveals that current state-of-the-art realtime voice agents lag behind non-realtime counterparts in multiparty conversation and both of them perform near-random performance on implicit and negative turn-taking, highlighting a critical limitation in current models.
Notably, these findings are robustly supported by our human evaluation, which exhibits a high correlation with the LLM judge.
Our main contributions are as follows:
\begin{enumerate}
    \item We introduce MP-Bench\footnote{https://github.com/atosystem/MP-Bench}, the first step toward benchmarking voice agents as active participants in multiparty conversations, spanning both understanding and behavior aspects.
    \item We evaluate 12 state-of-the-art voice agents, revealing that current models struggle significantly with appropriate multiparty interaction.
    \item Through ablation studies, we demonstrate that these shortcomings are non-trivial and stem directly from the inherent complexities of multiparty dynamics.
\end{enumerate}

\section{Related Works}
Recent voice agent benchmarks have targeted various dimensions, including paralinguistics~\cite{Ao_2024_SDEVAL,yang_2026_parass,liu_2025_emoreasoning}, full-duplex turn-taking~\cite{ge_2025_flexi,arora_2025_talkingturns,chang_2025_gametime,peng_2025_fdbench,lin_2025_fdbench_v1,lin_2025_fdbench_v15,lin_2026_fdbench_v2}, reasoning~\cite{cui_2025_voxeval,lin_2025_vera,fang_2025_s2sbench}, multi-turn coherence~\cite{du_2025_mtalk,gosai_2025_audiomultichallenge}, tool-use~\cite{jain_2026_voiceagentbench,lin_2026_fdbench_v3}, or a combination thereof~\cite{chen_2024_voicebench,liu_2025_vocalbench,yan_2025_urobench,li_2026_wavbench}. 
Existing multiparty benchmarks primarily assess passive understanding rather than active participation~\cite{inoue_2025_multimodal_multiparty,wang_2025_msubench,kwon_2025_m3slu}, and while concurrent work evaluates active turn-taking, it is strictly confined to text modality~\cite{bhagtani26speakstaysilentcontextaware,nama26_when2speak}. 
In contrast, our benchmark evaluates understanding, turn-taking, and response appropriateness directly from audio, forcing agents to jointly process acoustic and contextual cues, more closely mirroring real-world scenarios.

\section{MP-Bench Overview}
\label{subsec:data_structure}
MP-Bench consists of $927$ tasks created from two scenarios, \Scone~and \Sctwo, with each task belonging to one of two task families: \textit{understanding} tasks or \textit{behavioral} tasks.
The \Scone~scenario contributes tasks of both task families, whereas the \Sctwo~scenario contributes only behavioral-task tasks.
We provide the overview and evaluation description in this section and leave the construction process in Sec~\ref{sec:mpbench_construct}.

\begin{table}[t]
  \centering
  \footnotesize
  \setlength{\tabcolsep}{5pt}
  \begin{tabular}{lp{6cm}}
    \toprule
    Spk. & Content \\
    \midrule
    \multicolumn{2}{l}{\textit{\Scone~Scenario}} \\
    \midrule
    \textcolor{NavyBlue}{Mia} & \textcolor{NavyBlue}{Which continent has the highest number of countries in it?} \\
\textcolor{Purple}{Karen} & \textcolor{Purple}{I read somewhere that it's Africa.} \\
\textcolor{ForestGreen}{Jack} & \textcolor{ForestGreen}{Really? I feel like Europe has a ton of tiny countries.} \\
\textcolor{NavyBlue}{Mia} & \textcolor{NavyBlue}{That is a good point, Jack.} \\
\textcolor{Purple}{Karen} & \textcolor{Purple}{But Africa is massive.} \\
\textcolor{ForestGreen}{Jack} & \textcolor{ForestGreen}{Maybe, but Asia has a lot too. What is your final answer, Karen?} \\
\textcolor{Purple}{Karen} & \textcolor{Purple}{Sticking with Africa.} \\
\textcolor{NavyBlue}{Mia} & \textcolor{NavyBlue}{I am going with Europe.} \\
\textcolor{ForestGreen}{Jack} & \textcolor{ForestGreen}{Mia is probably wrong.} \\
\textcolor{NavyBlue}{Mia} & \textcolor{NavyBlue}{Rude! AI Assistant, please tell us the answer. And judge who is correct.} \\
\hline
\textcolor{black!75}{Agent} & \textcolor{black!75}{It's Africa, so Mia is correct.}~~~\textcolor{red}{\ding{55}} \\
\textcolor{black!75}{Agent} & \textcolor{black!75}{It's Africa, so Karen is correct.}~~~\textcolor{ForestGreen}{\ding{51}}\\
    \midrule
    \midrule
    \multicolumn{2}{l}{\textit{\Sctwo~Scenario}} \\
    \midrule
\textcolor{Purple}{Rose} & \textcolor{Purple}{Mia, Leo, are you two ready for a mental workout? I call it the Rival Response.} \\
\textcolor{ForestGreen}{Leo} & \textcolor{ForestGreen}{Always ready, Rose. How do we play?} \\
\textcolor{Purple}{Rose} & \textcolor{Purple}{Two teams. When a Group Leader says a word, the Group Follower from the other side has to respond randomly.} \\
\textcolor{NavyBlue}{Mia} & \textcolor{NavyBlue}{Sounds fun. How are we dividing up?} \\
\textcolor{Purple}{Rose} & \textcolor{Purple}{Mia, you take the helm of Group 1 as Leader, and Leo, you'll be the Follower for Group 1.} \\
\textcolor{ForestGreen}{Leo} & \textcolor{ForestGreen}{Solid copy. I'm following Mia.} \\
\textcolor{Purple}{Rose} & \textcolor{Purple}{I'll take the lead for Group 2, and the AI Assistant will be my Follower.} \\
\textcolor{NavyBlue}{Mia} & \textcolor{NavyBlue}{Okay, so if Rose speaks, Leo answers. If I speak, the AI answers.} \\
\textcolor{Purple}{Rose} & \textcolor{Purple}{Exactly. Focus up, everyone. Game start.} \\
\textcolor{Purple}{Rose} & \textcolor{Purple}{Penguin} \\
\textcolor{ForestGreen}{Leo} & \textcolor{ForestGreen}{Laser} \\
\textcolor{Purple}{Rose} & \textcolor{Purple}{Book} \\
\hline
\textcolor{black!75}{Agent} & \textcolor{black!75}{Balloon}~~~\textcolor{red}{\ding{55}} \\
\textcolor{black!75}{Agent} & \textcolor{black!75}{[silence]}~~~\textcolor{ForestGreen}{\ding{51}}\\
    \bottomrule
  \end{tabular}
  \caption{Examples behavior tasks generated from the two scenarios. In the top example, the agent must settle the discussion while on the bottom example the agent should remain silent since it is not its turn.
  We also provide appropriate and non-appropriate agent response for each example. More examples in Appendix~\ref{subsec:app_task_example}.}
\label{tab:scenario_example_instances}
\vspace{-12pt}
\end{table}

\subsection{Scenario}
\label{subsec:scenario_intro}
Both scenarios place the voice agent as the fourth participant in a four-way spoken interaction with three simulated humans.
Representative tasks from each scenario are shown in Table~\ref{tab:scenario_example_instances}.

\noindent
\textbf{\Scone.}
This scenario simulates a multiparty debate centered on a single question.
After the question is posed, the three human participants exchange differing viewpoints; the conversation concludes with one of them explicitly addressing the voice agent by name and inviting it to evaluate the arguments and declare which speaker is correct.
This setup is designed to evaluate whether the agent can comprehend a contested multiparty discussion and act on an explicit invitation to take its turn.

\noindent
\textbf{\Sctwo.}
The \Sctwo~scenario evaluates the agent's behavior within a structured, turn-based game.
A human speaker introduces the rules and leads a preliminary exchange that establishes the vocal characteristics of the four participants, who are split into two opposing teams; each team has a leader and a follower.
The core rule dictates that whenever a team leader speaks, the opposing team's follower must immediately respond.
These role assignments are introduced precisely to make turn-taking clear and deterministic: at any moment in the game, the voice agent's role within its team unambiguously specifies whether it is the participant expected to speak next.
The required spoken content depends on the game variant being played: \textit{Word Chain} (saying a random word), \textit{Increasing Number}, or \textit{Decreasing Number} (incrementing or decrementing the opposing leader's number).
We intentionally minimize the cognitive load of the required spoken content as the goal is to test multiparty turn-taking ability rather than complex reasoning and instruction following.

\subsection{Task}
\label{subsec:task_intro}
Each task in MP-Bench belongs to one of two families: \textit{understanding} tasks or \textit{behavioral} tasks.

\noindent
\textbf{Understanding tasks.}
In an understanding task, the audio clip ends with one speaker explicitly asking the voice agent an understanding question about the preceding conversation, and the agent's spoken answer is compared against a ground-truth answer.
These tasks are constructed from the \Scone~scenario and are designed to probe the agent's ability to comprehend multiparty conversations, including identifying speakers, tracking who said what, and resolving references across speakers.

\noindent
\textbf{Behavioral tasks.}
In a behavioral task, we assess the agent's action immediately following the audio segment: whether it appropriately takes a turn or remains silent, and, when it does speak, whether its response is contextually appropriate.
Each behavioral task carries a ground-truth label indicating whether the agent is expected to speak.
Behavioral tasks drawn from \Scone~target \textit{explicit} turn-taking, in which the agent is directly invited to respond; behavioral tasks drawn from \Sctwo~target \textit{implicit} and \textit{negative} turn-taking, in which the agent must infer from context whether to speak or remain silent and, when it does speak, comply with the established game rule.

\subsection{Evaluation}
\label{subsec:task_eval}

\subsubsection{Evaluation Pipeline}
Following the protocol of Full-Duplex-Bench~\cite{lin_2025_fdbench_v1}, we stream the input audio to the voice agents and transcribe their spoken output using the NeMo Parakeet streaming ASR model~\cite{rekesh_2023_parakeet_fastconformer}. 
We leverage ASR word-level timestamps to detect response onsets; an agent is flagged as having responded if any generated word starts after the conclusion of the input audio, successfully bypassing the need for an explicit voice activity detection (VAD) module. 
The generated spoken content is then assessed using an LLM-as-a-judge framework~\cite{chiang_2023_llmjudge} powered by GPT-5~\cite{openai_2025_gpt5}.

\subsubsection{Understanding Task Metrics}
\noindent
\textbf{Overall Accuracy.}
This metric measures whether the voice agent correctly answers the prompted understanding questions. 

\noindent
\textbf{Speaker Name Accuracy.}
Additionally, we report accuracy for a specific subset of tasks where the ground-truth answer corresponds to one of the speakers' names in the recording. 
We isolate this metric to enable direct comparison against a $33.3$\% random-guessing baseline, as every task involves exactly 3 human speakers.

\subsubsection{Behavioral Task Metrics}
\noindent
\textbf{Turn-Taking Accuracy.}
This metric evaluates the agent's ability to respond when required and remain silent otherwise. 
In the \Scone~scenario, the ground truth always requires a response (evaluating explicit turn-taking). 
In the \Sctwo~scenario, the ground-truth labels are evenly balanced between speaking and remaining silent, establishing a $50$\% random-guessing baseline.

\noindent
\textbf{Appropriateness Accuracy.}
Finally, we report an appropriateness accuracy where a sample is considered correct only if the agent exhibits both correct turn-taking behavior \textit{and} generates a valid response. 
For the \emph{\Scone}~scenario, a response is deemed valid if it shows logical consistency within the debate, regardless of factuality. 
For the \emph{\Sctwo}~scenario, validity is defined as compliance with the assigned game rules.

\subsubsection{Response Latency}
For each task, we report the mean \textit{response latency} for realtime voice agents. 
This is defined as the duration between the end of the input audio (i.e., the conclusion of the last input word) and the start of the agent's first generated word. 
These timings are extracted from the ASR timestamps results same as turn-taking detection.
\section{MP-Bench Construction}
\label{sec:mpbench_construct}
\begin{figure*}[t]
  \includegraphics[width=\linewidth,trim={0.7cm 0.3cm 0.7cm 0.3cm},clip]{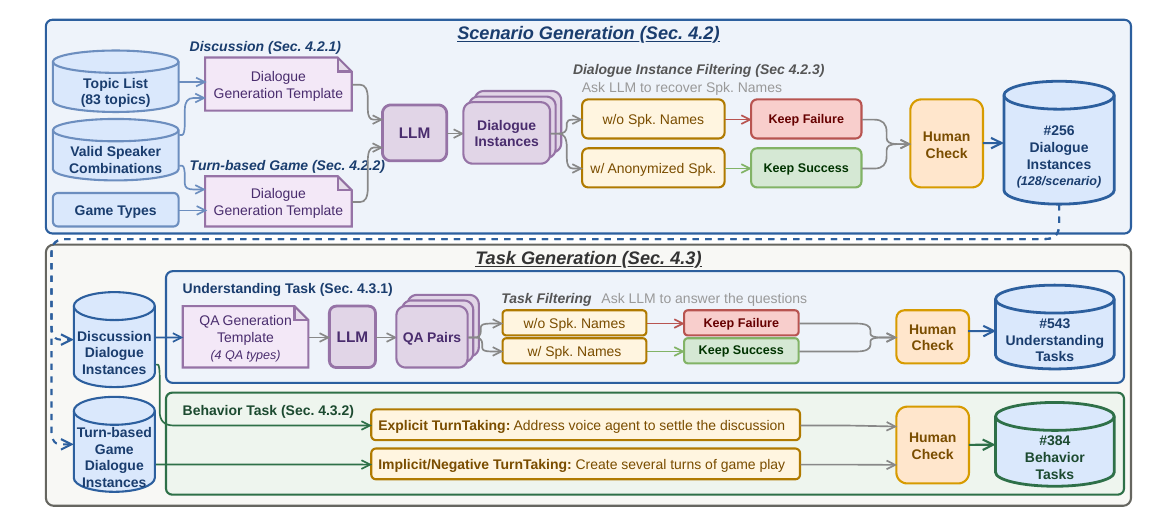}
  \caption {The overall pipeline for creating MP-Bench. We first generate dialogue contents from the two scenarios and then generate \emph{Understanding} and \emph{Behavior} tasks for each generated dialogue instance.}
  \label{fig:benchmark-gen-pipeline}
\end{figure*}

The construction of MP-Bench is guided by two core principles: (1) tasks must be objective and automatically evaluable, and (2) tasks must necessitate auditory cues, ensuring they cannot be trivially solved by text-only LLMs using transcripts alone. 
To satisfy these principles, we require precise control over the spoken dialogue. 
Therefore, we employ LLMs to generate the conversation content and ElevenLabs\footnote{https://elevenlabs.io}' conversation API to synthesize high-quality conversational audio.

Fig.~\ref{fig:benchmark-gen-pipeline} shows our benchmark generation pipeline. 
We begin by curating a pool of speaker combinations (Sec.~\ref{subsec:spk_attr}) and defining the two \textbf{scenarios} (Sec.~\ref{subsec:scenario_intro}, \ref{subsec:scenario_generation}). Leveraging these components, we prompt an LLM to generate diverse \textbf{dialogue instances}. 
Finally, for each instance, we construct specific evaluation \textbf{tasks} (Sec.~\ref{subsec:task_intro}, \ref{subsec:task_gen}).
To ensure high quality, all tasks undergo rigorous human verification before being finalized for the benchmark. 
Detailed statistics in Appendix~\ref{subsec:app_mpbench_statistics}.

\subsection{Speaker Attribution}
\label{subsec:spk_attr}
We evaluate voice agents in a four-participant setting to balance multiparty complexity with manageable audio lengths. 
This choice is deliberate: 3 speakers (2 pre-recorded plus the agent itself) simplifies the task to a near-dyadic baseline easily solved by standard ASR and LLM pipelines, while more than 4 introduces long-context memory failures that confound evaluation. 4 participants captures true multiparty dynamics without triggering context-length bottlenecks, though varying speaker counts remains a natural direction for future work as long-context robustness improves.

To ensure objective evaluation, we carefully curate voice sets with highly distinct acoustic characteristics, yielding high speaker separability (Appendix~\ref{subsec:app_mpbench_audio_quality}). 
These curated sets are designated as ``valid speaker combinations'' and used during dialogue generation (See Appendix~\ref{subsec:app_spk_attribution}). 
To evaluate benchmark difficulty under harder acoustic conditions, we also conduct an ablation using deliberately similar voices across all pre-recorded audio. Further details are provided in Appendix~\ref{sec:app_similar_voices}.

\subsection{Scenario Generation}
\label{subsec:scenario_generation}

\subsubsection{\Scone~Scenario}
We curate a list of 83 topics (Appendix~\ref{subsec:app_topics}) and design a corresponding prompt template for the LLM. 
To generate each dialogue instance, we randomly sample a topic from the list alongside a valid speaker combination to populate the template, which is then fed into an LLM for generation. 
To maximize diversity and prevent duplication, we maintain a strict record of all generated topic-speaker tuples. 

Furthermore, the LLM generates two additional types of content-related metadata for each spoken turn to facilitate subsequent question generation: addressees and coreferences. 
Addressees indicate whom the speaker is addressing during a given utterance; notably, this is not necessarily the subsequent speaker. 
Coreferences are pronouns or noun phrases in utterances that refer back to the conversational participants or previously discussed topics.

\subsubsection{\Sctwo~Scenario}
Similar to Scenario 1, we design a prompt template and randomly sample valid speaker combinations to guide the LLM in generating the dialogue transcripts.
For this scenario, we do not ask LLM to generate the coreferences and the addressee as this scenario are not suitable due to the game setup.

\subsubsection{Dialogue Instance Filtering}
For both scenarios, to ensure that the generated dialogues cannot be trivially solved by a text-only LLM, we must guarantee that speaker identities cannot be fully recovered solely from the transcript. Conversely, to make the task solvable via the audio modality, every participant's name must be explicitly mentioned at least once during the conversation; otherwise, it would be impossible for the voice agent to map the acoustic profiles to their respective identities. 

In practice, we apply a rule-based filter to verify that each participant is addressed or mentioned at least once. 
Furthermore, we employ an LLM to validate the generated dialogue under two distinct prompting conditions: (a) providing only the raw spoken content without any speaker information, and (b) providing the spoken content paired with anonymized speaker identifiers~(Speaker1, Speaker2...) for each turn.
We then filter the generated dialogues, retaining only those instances that the LLM successfully solves all speaker names for each spoken turn under condition (b) but fails to solve under condition (a). 
Notably, condition (b) effectively simulates the theoretical upper bound of a perfect cascaded voice agent equipped with flawless ASR and speaker diarization systems.

\begin{table}[t]
  \centering
  \small
  \setlength{\tabcolsep}{5pt}
  \begin{tabular}{lccp{4cm}}
    \toprule
    \# & Inquiry & Based on & Example \\
    \midrule
    1 & Spk. & Cont. & \textit{What's the speaker's name who think Paris is a capital?} \\
    2 & Cont. & Spk. & \textit{What is Jack certain about in the conversation?} \\
    3 & Cont. & Cont. & \textit{What's the claim of the speaker who started the question?} \\
    4 & Spk. & Spk. & \textit{What's the speaker's name that disagree with Jack's answer?} \\
    \bottomrule
  \end{tabular}
  \caption{Examples for the 4 types of QA in our benchmark. ``Spk.'' and ``Cont.'' indicates speaker attributes and content-related information respectively. }
  \label{tab:qa_tasks_examples}
\end{table}

\subsection{Task Generation}
\label{subsec:task_gen}
\subsubsection{Understanding Tasks}
\label{subsec:understand_task_gen}
We employ \Scone~dialogue instances to create question-answering tasks.
To generate QA pairs systematically, we first categorize the information in the generated dialogue instances into two categories: speaker-related attributes (e.g., name, gender, voice characteristics) and content-related details (e.g., spoken utterances, addressees, and coreferences).
Based on these two types of information, we construct four distinct categories of questions~(Tab.~\ref{tab:qa_tasks_examples}).

\begin{table*}[t]
  \centering
  \small
  \setlength{\tabcolsep}{3pt}
  \resizebox{\textwidth}{!}{%
  \begin{tabular}{lccc@{\hspace{4pt}}ccc@{\hspace{4pt}}ccc}
    \toprule
    \multirow{2}{*}{\textbf{Model}}
     & \multicolumn{3}{c}{\textbf{Understanding: \Scone}}
     & \multicolumn{3}{c}{\textbf{Behavior: \Scone}}
     & \multicolumn{3}{c}{\textbf{Behavior: Turn-based}} \\
     \cmidrule(lr){2-4} \cmidrule(lr){5-7} \cmidrule(lr){8-10}
     & Overall Acc. & Spk.\ Name Acc. & Lat.\ (s) & TT Acc. & App Acc. & Lat.\ (s) & TT Acc. & App Acc. & Lat.\ (s) \\
    \midrule
    \multicolumn{10}{l}{\cellcolor{gray!10}{\textit{Text LLMs}}} \\
    \textlm{Gemini-3.1-Pro (ASR-only)}  & \textlm{\phantom{0}4.79} & \textlm{27.27} & \textlm{---} & \textlm{100}   & \textlm{75.00} & \textlm{---} & \textlm{49.22} & \textlm{49.22} & \textlm{---} \\
    \textlm{Gemini-3.1-Pro (SpkD-ASR)}  & \textlm{87.48} & \textlm{93.94} & \textlm{---} & \textlm{100}   & \textlm{89.06} & \textlm{---} & \textlm{98.44} & \textlm{98.44} & \textlm{---} \\
    \midrule
    \multicolumn{10}{l}{\cellcolor{gray!10}{\textit{Non-realtime Voice Agents}}} \\
    Qwen3-Omni                 & 34.44 & 36.11 & --- & 100   & 60.16 & --- & \textbf{52.73} & 46.87 & --- \\
    Gemini-3.1-Pro             & \textbf{73.30} & \textbf{78.28} & --- & 100   & \textbf{90.63} & --- & \textbf{52.73} & \textbf{51.95} & --- \\
    \midrule
    \multicolumn{10}{l}{\cellcolor{gray!10}{\textit{Realtime Voice Agents}}} \\
    Freeze-Omni                & \phantom{0}0.18 & \phantom{0}0.00 & 1.8 & 17.71 & \phantom{0}0.00 & 1.0 & 49.61 & 48.44 & \textbf{0.6} \\
    Moshi                      & \phantom{0}1.66 & \phantom{0}0.00 & 0.9 & 75.00 & \phantom{0}1.56 & \textbf{0.5} & \textbf{52.73} & 28.12 & 1.5 \\
    PersonaPlex                & \phantom{0}3.14 & \phantom{0}7.32 & \textbf{0.6} & 96.09 & 0.00 & \textbf{0.5} & 51.17 & \phantom{0}4.69 & 0.7 \\
    Ultravox                   & 21.36 & 25.00 & 4.5 & 99.22 & 39.06 & 3.6 & 46.09 & 31.25 & 4.0 \\
    GLM-4-Voice                & 11.60 & 12.63 & 2.4 & \textbf{100.00} & 34.38 & 2.3 & 49.14 & 45.35 & 2.6 \\
    Covo-Audio                 & \textbf{32.60} & \textbf{37.37} & 1.9 & 99.22 & 42.19 & 1.7 & 48.83 & 48.44 & 1.3 \\
    GPT-Realtime               & 14.00 & 16.67 & 5.9 & \textbf{100.00} & 53.13 & 6.0 & 48.44 & 28.13 & 6.7 \\
    DuplexCascade              & 6.45 & 9.23 & 2.7 & \textbf{100.00} & 3.91 & 2.1 & 48.63 & \textbf{48.63} & 17.2 \\
    Gemini-2.5 Live            & 10.58 & 11.87 & 8.5 & 78.13 & 36.72 & 6.5 & 48.00 & 20.00 & 8.6 \\
    Gemini-3.1 Live            & \phantom{0}4.90 & \phantom{0}5.81 & 8.7 & 99.22 & \textbf{57.03} & 7.4 & 52.57 & 38.99 & 8.2 \\
    \bottomrule
  \end{tabular}%
  }
  \caption{%
  Main MP-Bench results. TT. Acc.\ and Lat.\ denote turn-taking accuracy (\%) and latency (seconds), respectively. App Acc. indicates appropriateness accuracy. Random-guessing baselines are 33.3\% for Spk.\ Name Acc.\ and 50\% for Turn-Based TT. Acc. For Discussion TT. the ground truth is always respond, please read jointly with the balanced Turn-based metrics. Latency is measured exclusively for real-time voice agents and averaged over non-silent responses.
  }
  \label{tab:main_bench_results}
\end{table*}

We explicitly prompt an LLM to generate QA pairs corresponding to one of these four defined types.
Subsequently, we apply an LLM-based filtering mechanism, similar to the protocol employed during the scenario generation.
Specifically, we evaluate the generated questions under two prompting conditions: (a) providing a text-only LLM with the raw spoken transcript, and (b) providing the LLM with the speaker names. 
We retain only those QA pairs that the LLM successfully resolves under condition (b) but fails under condition (a). 
This ensures that the questions are inherently logical and solvable with full context, yet cannot be trivially answered using solely the text transcript, thereby necessitating true multiparty spoken conversation comprehension.

\subsubsection{Behavior Tasks}
\label{subsec:behave_task_gen}
For the \textit{\Scone} scenario dialogue instances, we append an additional spoken turn at the end of the dialogue instance that explicitly addresses the voice agent~(Table~\ref{tab:scenario_example_instances}), asking which participant it agrees with. 
For the \textit{\Sctwo} scenario, we generate two distinct sets of game-playing turns and append them following the dialogue instances~(i.e., the rules introduction and team assignment).

\section{Benchmark Results}
\subsection{Evaluated Models}
\label{sec:evaluated_models}

\noindent\textbf{Realtime Audio Models.} 
We evaluate 10 systems designed for realtime spoken interaction, comprising proprietary full-duplex APIs (GPT-Realtime~\cite{openai_2025_gptrealtime}, Gemini-2.5 Live~\cite{deepmind_2025_gemini25}\footnote{We refer to \texttt{gemini-2.5-native-dialogue}.}, Gemini-3.1 Live~\cite{google_2025_geminiliveapi}), open-source full-duplex streaming models (Moshi~\cite{defossez_2024_moshi}, PersonaPlex~\cite{roy_2026_personaplex}, Freeze-Omni~\cite{wang_2024_freezeomni}, and Ultravox~\cite{fixie_2024_ultravox}), and half-duplex end-to-end speech LMs (GLM-4-Voice~\cite{zeng_2024_glm4voice} and Covo-Audio~\cite{wang_2026_covoaudio}).

\noindent\textbf{Non-real-time Audio Models.} Unsuited for low-latency interaction, this group comprises non-streaming omni models. 
We evaluate two systems: the non-streaming Qwen3-Omni~\cite{xu_2025_qwen3omni} and Gemini-3.1-Pro~\cite{deepmind_2025_gemini3}, which process full audio inputs to produce text.
These models serve as a strong top line for audio-native multiparty comprehension when the realtime interaction constraint is relaxed.

\noindent\textbf{Text LLMs.} 
As references, we evaluate Gemini-3.1-Pro~\cite{deepmind_2025_gemini3} under two configurations.
In the \textit{ASR-only} setting, the model receives the ground truth ASR transcript of the input audio without any speaker label; this captures the performance ceiling of a cascaded system that lacks speaker diarization.
In the \textit{SpkD-ASR} setting, the model additionally receives the ground-truth anonymized speaker label for every turn, emulating the upper bound of a cascaded system equipped with perfect speaker diarization.
For both settings, the response is generated as text and compared to the reference answer directly.

\subsection{Main Results}
Table~\ref{tab:main_bench_results} summarizes the MP-Bench results.
We highlight several notable trends in the following.

\noindent
\textbf{Behavioral Tasks are Less Challenging than Understanding under Explicit Multiparty Turn-Taking.}
We observe divergent performance trends between behavioral and understanding tasks within the \textit{\Scone} scenario. 
Across nearly all settings, models achieve significantly higher scores on behavioral tasks than on understanding. 
This discrepancy arises because understanding tasks require the model to accurately understand fine-grained acoustic and contextual relationships. 
In contrast, behavioral tasks allow powerful LLMs to rely on their strong linguistic priors to generate reasonable responses, even when they fail to fully resolve the multiparty dynamics. 
Notably, the performance gap between the text LLM ASR-Only and SpkD-ASR settings shrinks dramatically from $66.67$\% (93.94 - 27.27) on understanding to just $14.06$\% (89.06 - 75.00) on behavioral appropriateness accuracy. 
This confirms that highly capable LLMs can effectively infer or hallucinate reasonable conversational continuations even in the absence of explicit speaker identification.

\noindent
\textbf{Voice Agents Struggle with Implicit and Negative Turn-Taking Relative to Explicit Cues.}
Comparing turn-taking accuracy between the \textit{\Sctwo} and \textit{\Scone} scenarios reveals a substantial performance gap. 
Nearly all voice agents perform at or below the random-guessing baseline in the \Sctwo, whereas multiple realtime agents achieve over $90$\% accuracy in the \Scone~scenario. 
In contrast, the text LLM baseline equipped with SpkD-ASR achieves near-perfect turn-taking accuracy across both. 
These results indicate that a primary challenge for current voice agents lies in identifying implicit turns and appropriately remaining silent when they are not addressed.
Furthermore, the near-perfect performance of the SpkD-ASR baseline confirms that the task is fundamentally solvable, provided the model can accurately extract and leverage speaker-attributed information during prediction.

\noindent
\textbf{Non-Realtime Agents Outperform Realtime Models, Particularly on Understanding.} 
In understanding tasks, non-realtime agents (led by Gemini-3.1-Pro) significantly outperform their realtime counterparts. 
Even the best realtime system, Covo-Audio, lags far behind Gemini-3.1-Pro in both overall and speaker name accuracy.  
Notably, most realtime models fail to clear the $33.3$\% random baseline for speaker identification, whereas both non-realtime systems exceed this threshold. 
Within the realtime group, Covo-Audio stands out as the lone strong performer; Moshi and Freeze-Omni produce mostly generic backchannels at the bottom. 
This gap indicates that low-latency architectural trade-offs, such as streaming decoding, severely penalize multiparty understanding.

In \textit{\Scone} behavioral tasks, the realtime versus non-realtime performance gap persists but narrows. However, in \textit{Turn-Based Games}, this delta vanishes entirely, with all models performing at near-random levels. 
These findings suggest that while multiparty comprehension can be improved by sacrificing latency, the inability to navigate implicit and negative turn-taking is a fundamental deficit orthogonal to latency constraints.

\section{Ablation Results}
\subsection{Disentangling Game-Rule Following from Multiparty Capabilities}
\label{sec:ablation_single_speaker}

\begin{table}[t]
  \centering
  \small
  \setlength{\tabcolsep}{4pt}
  \begin{tabular}{lccc}
    \toprule
    \textbf{Model} & \textbf{Precision} & \textbf{Recall} & \textbf{TT Acc.} \\
    \midrule
    GPT-Realtime    & 78.2 & 70.0 & 76.0 \\
    Gemini-2.5 Live & 64.0 & 90.0 & 68.0 \\
    \bottomrule
  \end{tabular}
  \caption{Single-speaker Turn-Based game ablation~(\%). $50$ ElevenLabs-synthesized dialogues featuring a single human speaker who explains the rules and plays the game directly with the agent. Both models are clearly above the $50$\% random baseline, unlike the multiparty Turn-Based game, where every realtime model is within $\pm\!3$ points of chance.}
  \label{tab:ablation_single_spk}
\end{table}

The multiparty \textit{Turn-Based Games} scenario demands two concurrent capabilities from the agent: adhering to game rules to generate appropriate content, and utilizing multi-party context to determine when it is appropriate to speak. 
To test whether the observed failures stem from complex game rules, we construct a \emph{single-speaker} variant.
In this setup, one human speaker introduces the rules, assigns roles, and plays directly with the agent, while the core mechanics (Word Chain, Increasing Number, Decreasing Number) remain identical to the multiparty version.

Evaluating $50$ ElevenLabs-synthesized single-speaker dialogues (Table~\ref{tab:ablation_single_spk}) reveals that 
both GPT-Realtime ($76.0\%$ TT Acc.) and Gemini-2.5 Live ($68.0\%$) perform significantly above the $50\%$ random baseline (exact binomial, $p = 3.1 \times 10^{-4}$ and $p = 0.015$; $95\%$ intervals $[62.6, 85.7]$ and $[54.2, 79.2]$, both excluding chance), and their appropriateness accuracy ($74.0\%$ and $63.0\%$) far exceeds their multiparty counterparts ($28.13\%$ and $20.00\%$).
In stark contrast, their performance on the multiparty equivalent hovers within $\pm 3$ points of chance (Table~\ref{tab:main_bench_results}). 
These results demonstrate that the multiparty failure is not a failure of game-rule instruction following. We note, however, that the single-speaker variant also removes the speaker-identification burden, so this ablation alone is not aim to separate turn-taking from speaker attribution. 
That separation is instead provided by the text-LLM references in Table~\ref{tab:main_bench_results}. Turn-Based turn-taking is at chance given transcripts alone (ASR-only, 49.22) but near-ceiling with oracle speaker labels (SpkD-ASR, 98.44), indicating the deficit lies in recovering speaker attribution from audio rather than in reasoning about turns.

\subsection{Multiparty-aware Prompting}
\label{sec:ablation_prompt}
Given the generally poor performance of voice agents in multiparty settings, we investigate whether this limitation can be trivially resolved through prompting. 
In our primary evaluation (Sec.~\ref{sec:evaluated_models}), we employ the \textit{default} system prompt for all applicable voice agents. 
For this ablation, we design \textit{multiparty-aware} system prompts that explicitly inform the agent that ``there may be multiple speakers in the conversation.'' 
Both prompts are provided verbatim in Appendix~\ref{sec:app_realtime_prompts}.

\begin{table}[t]
  \centering
  \small
  \setlength{\tabcolsep}{3pt}
  \renewcommand{\arraystretch}{0.85}
  \resizebox{\columnwidth}{!}{%
  \begin{tabular}{lcccc}
    \toprule
    \multirow{2}{*}{\textbf{Model}}
     & \textbf{Und.}
     & \textbf{Disc.}
     & \multicolumn{2}{c}{\textbf{Turn-Based}} \\
     \cmidrule(lr){2-2} \cmidrule(lr){3-3} \cmidrule(lr){4-5}
     & OA & App & TT & App \\
    \midrule
    GPT-Realtime~(default)        & 14.00 & 53.13 & 48.44 & 28.13 \\
    GPT-Realtime~(MP-aware)       & 16.39 & 50.02 & 49.61 & 26.17 \\
    \midrule
    Gemini-2.5 Live~(default)     & 10.58 & 36.72 & 48.00 & 20.00 \\
    Gemini-2.5 Live~(MP-aware)    & 13.63 & 85.16 & 48.39 & 18.95 \\
    \midrule
    Gemini-3.1 Live~(default)     & \phantom{0}4.90 & 57.03 & 52.57 & 38.99 \\
    Gemini-3.1 Live~(MP-aware)    & \phantom{0}7.82 & 57.03 & 47.41 & 25.96 \\
    \midrule
    Ultravox~(default)            & 21.36 & 39.06 & 46.09 & 31.25 \\
    Ultravox~(MP-aware)           & 19.44 & 39.06 & 51.95 & 12.50 \\
    \midrule
    PersonaPlex~(default)         & \phantom{0}3.14 & \phantom{0}0.00 & 51.17 & \phantom{0}4.69 \\
    PersonaPlex~(MP-aware)        & \phantom{0}3.13 & \phantom{0}0.00 & 51.95 & \phantom{0}5.47 \\
    \bottomrule
  \end{tabular}%
  }
  \caption{Effect of inference-time system prompts~(default vs.\ multi-party-aware) on MP-Bench scores~(\%). Abbreviations: OA (Overall Accuracy), App (Appropriateness Accuracy), and TT (Turn-Taking Accuracy). Und. and Disc. indicates understanding and behavior tasks from \emph{\Scone}~scenario respectively.}
  \label{tab:ablation_prompt}
\end{table}

Table~\ref{tab:ablation_prompt} reports the comparison for the five real-time models evaluated under both prompts; the default-prompt rows match those in Table~\ref{tab:main_bench_results}. 
The most significant improvements are concentrated in \textit{\Scone} scenario, where explicitly reminding agents of multiple speakers yields substantive gains for several systems. 
In contrast, overall performance either stagnates or degrades in the \textit{Turn-Based Games} scenario. 
Since TT Acc.\ remains near the 50\% random baseline under both prompts, the drop in Appropriateness Accuracy cannot be attributed to a change in turn-taking behavior. 
Instead, it reflects weaker rule compliance in the responses agents produce when they do choose to speak: under the multiparty-aware prompt, the content of their utterances departs more often from the game rules.

\section{Human Evaluation}

In addition to LLM Judge results, we also provide a small scale of human evaluation to make sure the alignment between human and LLM Judge is high.
We recruited human annotators to independently score a subset of 280 evaluation data entries equally sampled across all 14 settings from \emph{behavior} tasks from  \emph{\scone} scenario, given that the appropriateness accuracy of this task is the most subjective compared to other metrics that all have a specific ground truth. 
Human annotators are given the same transcripts and also an instruction  simplified from the LLM judge prompt.
More details of human evaluation setup is provided in Appendix~\ref{sec:app_human_eval}.

We measure Krippendorff's $\alpha$ and Cohen's $\kappa$. 
The former is an inter-annotator agreement coefficient computed over a unified rater pool consisting of the three human annotators.
The latter is the agreement between the LLM judge's predicted label and the per-item majority vote of the three human annotators.
The results show reasonably strong alignment
across the metrics (Krippendorff's $\alpha$ = 0.745, Cohen's $\kappa$ = 0.646), suggesting that the LLM-as-a-judge evaluation is reliable and substantially aligned with human evaluation.

\section{Conclusion}

We present MP-Bench, the first step toward benchmarking voice agents as active participants within multiparty interactions. 
Our evaluations reveal a stark capability gap: while non-realtime models demonstrate robust comprehension, their performance collapses to near-random levels during active participation. 
Meanwhile, realtime agents struggle broadly across both understanding and behavioral metrics. 
Crucially, our ablation studies confirm that these failures stem fundamentally from the inherent complexities of multiparty dynamics rather than task-specific nuances or general instruction-following deficits.
Furthermore, we demonstrate that these severe shortcomings cannot be circumvented via trivial interventions like explicit system prompting. 
Ultimately, MP-Bench exposes multiparty awareness as a foundational missing capability, establishing a clear mandate for the next generation of conversational AI.

\section{Limitation}

\paragraph{Scope.} 
MP-Bench is a controlled first step rather than a full test of open-ended participation. Because each task evaluates a single decision point at the end of a pre-recorded clip, we do not assess multi-turn engagement, barge-in, or backchanneling. Additionally, an agent's response is only registered if ASR detects a word onset after the input audio concludes. Any utterances by full-duplex models during the clip are ignored by our metrics. Consequently, MP-Bench evaluates voice agents under a conservative setting where competence is a necessary, but not sufficient, condition for natural multiparty interaction.

\paragraph{Synthetic Audio.} 
Controlled speech synthesis enables an objective ground truth for turn-taking that spontaneous, real-world recordings lack. However, this introduces a domain gap (Table~\ref{tab:audio_quality_comparison}). Our benchmark features no overlapping speech ($0\%$ vs.\ AMI's $24.5\%$), background noise, reverberation, long pauses, or disfluencies, though inter-turn timing for active transitions remains naturalistic ($0.40$\,s vs.\ $0.43$\,s). Because existing agents already struggle on these cleaned inputs, our simplified setup provides a conservative lower bound on model limitations. Incorporating real or actor-voiced recordings remains an important direction for future work.

\bibliography{custom}

\appendix

\section{Details of MP-Bench}
\label{sec:app_benchmark_detail}
\begin{figure}[h]
\includegraphics[width=\linewidth,trim={0.4cm 0.4cm 0.4cm 0.4cm},clip]{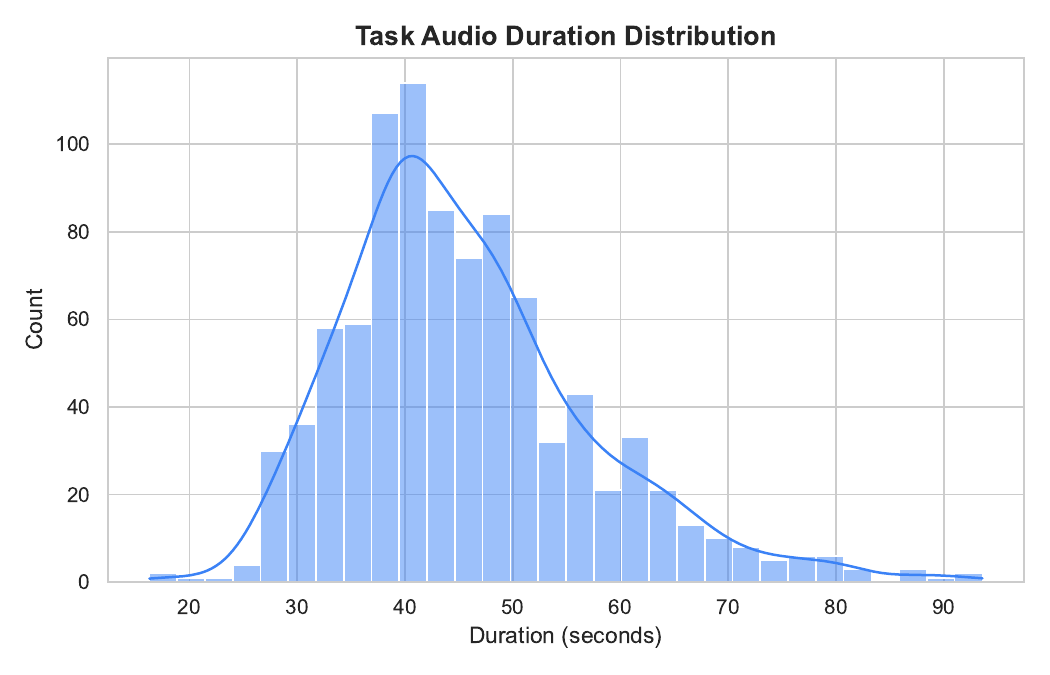}
  \caption {Task audio duration distribution.}
  \label{fig:app_audio_dur_dist}
\end{figure}
\begin{figure}[h]
\includegraphics[width=\linewidth,trim={0.4cm 0.4cm 0.4cm 0.4cm},clip]{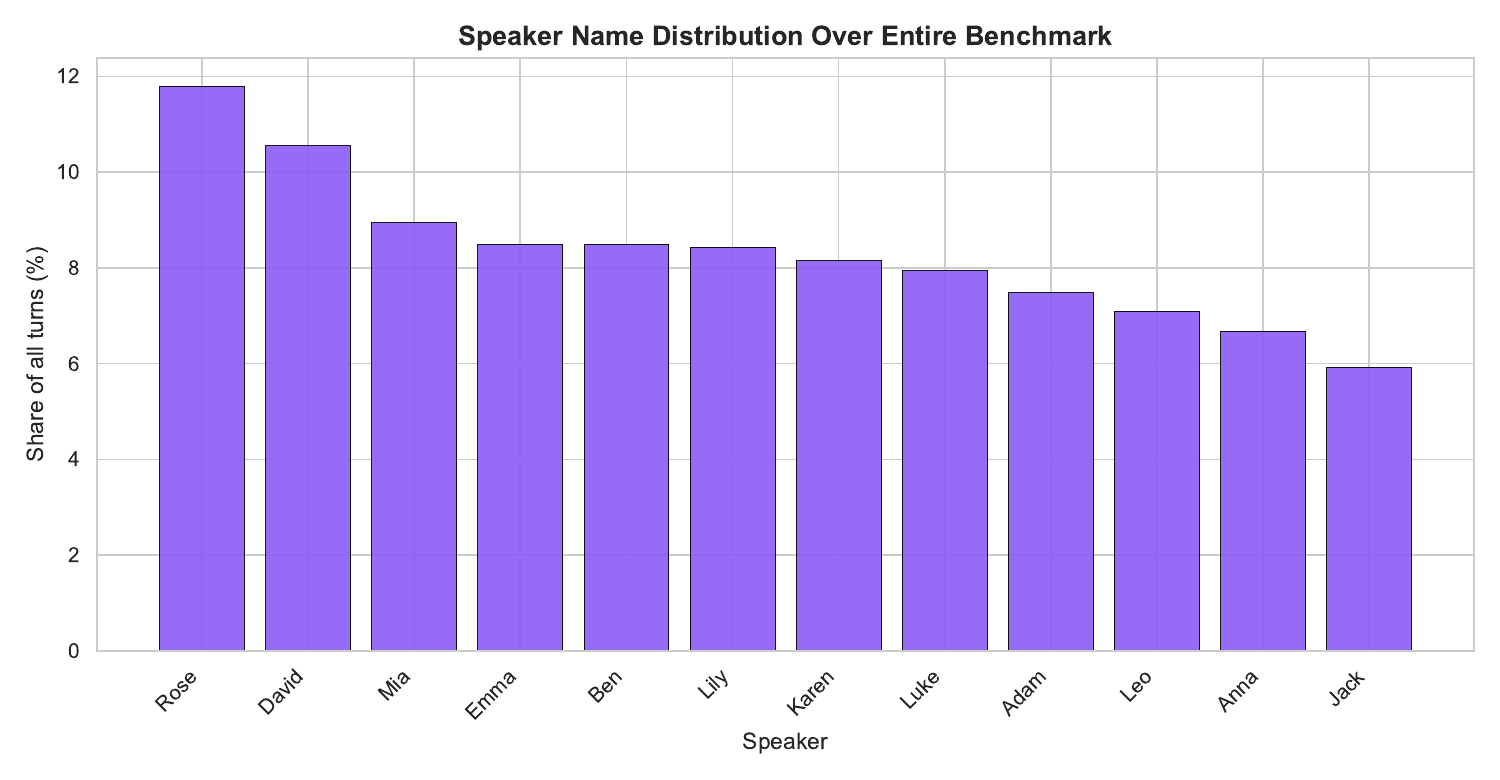}
  \caption {Overall Speaker Name Distribution.}
  \label{fig:app_speaker_dist}
\end{figure}
\begin{figure}[h]
\includegraphics[width=\linewidth,trim={0.4cm 0.4cm 0.4cm 0.4cm},clip]{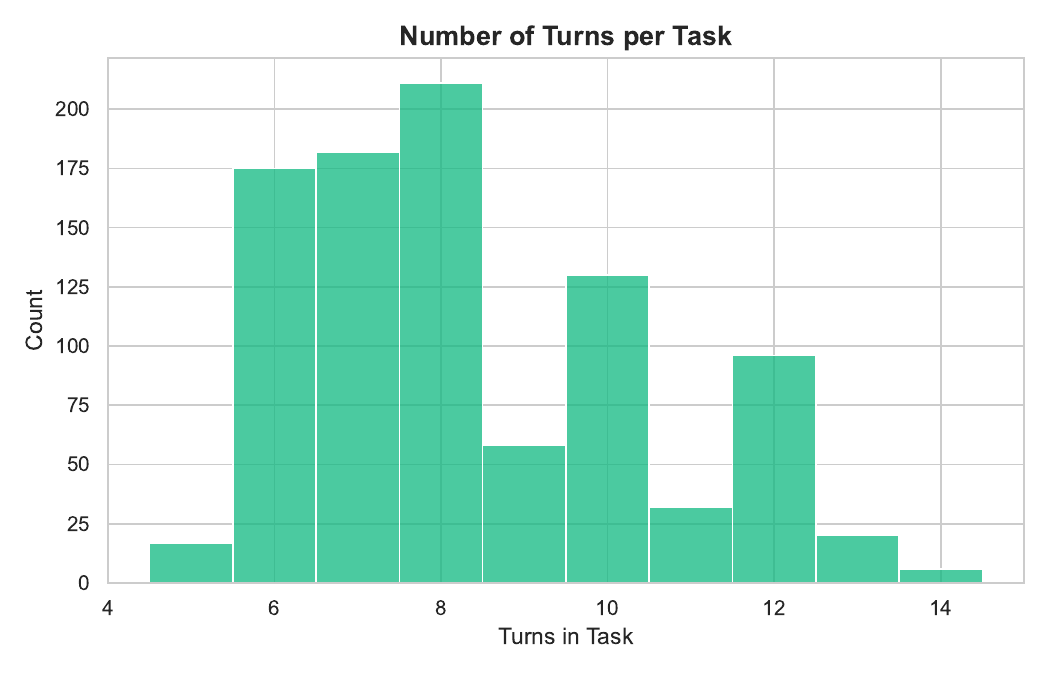}
  \caption {Number of Turns per Task.}
  \label{fig:app_turns_per_task}
\end{figure}
\begin{figure}[h]
\includegraphics[width=\linewidth,trim={0.4cm 0.4cm 0.4cm 0.4cm},clip]{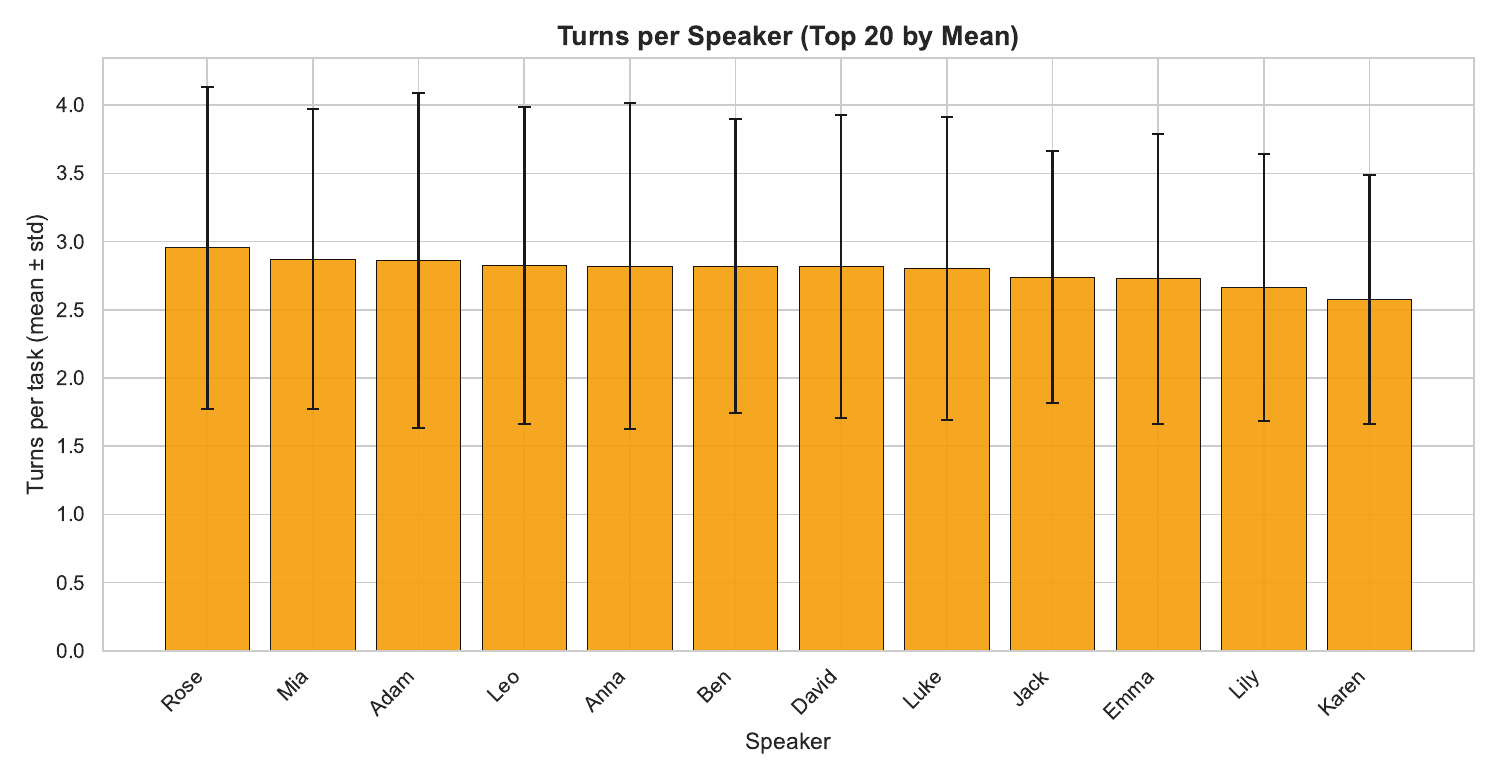}
  \caption {Number of Turns per Speaker for each Task.}
\label{fig:app_turns_per_speaker}
\end{figure}
\subsection{Statistics}
\label{subsec:app_mpbench_statistics}
We present the core structural statistics of the MP-Bench dataset. 
On average, each task recording lasts $45.6 \pm 11.5$ seconds and comprises $8.4 \pm 2.1$ conversational turns. 
Within individual tasks, each participant contributes an average of $2.8 \pm 1.1$ turns. 
Across the entire dataset, speaker participation is well-balanced, with each unique speaker accounting for a mean of $647.3 \pm 124.7$ total turns, representing approximately an $8.3\% \pm 1.6\%$ share of the overall dialogue. 
Comprehensive visualizations of these metrics—including the audio duration distribution (Fig.~\ref{fig:app_audio_dur_dist}), overall speaker distribution (Fig.~\ref{fig:app_speaker_dist}), and per-task turn frequencies (Figs.~\ref{fig:app_turns_per_task} and \ref{fig:app_turns_per_speaker})—are provided for further detail.

\begin{table*}[t]
\centering
\begin{tabular}{lcc}
\toprule
\textbf{Metric} & \textbf{MP-Bench (ours)} & \textbf{AMI (real)} \\
\midrule
UTMOS (quality, 1--5) & \textbf{3.88 $\pm$ 0.556} (4.05) & 2.83 $\pm$ 0.725 (2.97) \\
ASR WER (\%) & \textbf{6.6 $\pm$ 6.0 (5.0)} & 21.7 $\pm$ 5.4 (21.5) \\
Intra-speaker similarity & 0.571 $\pm$ 0.022 (0.567) & 0.387 $\pm$ 0.067 (0.391) \\
Inter-speaker similarity & 0.159 $\pm$ 0.024 (0.153) & 0.088 $\pm$ 0.031 (0.083) \\
Conversational turn gap duration (s) & 0.412 $\pm$ 0.235 (0.405) & 1.796 $\pm$ 2.634 (1.170) \\
Conversational turn gap duration, gaps $<$ 1 s (s) & 0.401 $\pm$ 0.177 (0.405) & 0.434 $\pm$ 0.278 (0.400) \\
Overlap time (\%) & 0.0 $\pm$ 0.0 (0.0) & 24.5 $\pm$ 10.4 (23.3) \\
Mean turn duration (s) & 2.91 $\pm$ 0.937 (2.81) & 2.65 $\pm$ 0.480 (2.62) \\
\bottomrule
\end{tabular}
\caption{Comparison of metrics between MP-Bench and AMI.}
\label{tab:audio_quality_comparison}
\end{table*}

\subsection{Speaker Attribution}
\label{subsec:app_spk_attribution}
The initial step in constructing the benchmark is to establish the speaker profiles for the dialogue. Each dialogue instance in our benchmark comprises 3 simulated human speakers, resulting in a four-participant conversation when the voice agent is included. 
To minimize subjectivity in our evaluation tasks, we ensure that the three speakers sampled for any given instance possess highly distinct vocal characteristics. 
Additionally, we provide an explicit textual description of each speaker's acoustic profile. 

In practice, we curate 12 distinct voices from the ElevenLabs library and categorize them along two primary axes: 1) gender (male or female) and 2) pitch (high, medium, or low).
To maximize the difference between the speakers within the same conversation, we do not allow speakers with high medium and low pitch of the same gender to be selected, as it might be tricky for evaluation.
As a result, to sample a valid speaker combination, we either select 2 male 1 female or 1 male 2 female speakers.
For each gender, we select 1 or 2 random distinct pitch. 
For example, (male high pitched, male medium pitched, female medium pitch) is a valid combination.
Other than the two primary axes, we also include several special description for each speaker voice.
By concatenating all of them, we create voice characteristic description for each voice.
This description will be used as a information when we create tasks in the following sections.

\subsection{Audio Quality Analysis}
\label{subsec:app_mpbench_audio_quality}
To provide readers with audio quality of MP-Bench, we calculated several statistics of our generated audio and the comparison with AMI corpus (as a real Multiparty recording reference).
We read Table~\ref{tab:audio_quality_comparison} in two ways, and we are careful not to overclaim:

\begin{itemize}
    \item \textbf{Cleanliness / intelligibility.} The audio is clean and highly intelligible: WER is 6.6\% vs.\ 21.7\% on real AMI recordings, and UTMOS is high. We do \textit{not} interpret UTMOS $>$ AMI as our audio being ``more natural than real conversation''. Clean synthesized speech is \textit{expected} to score above spontaneous, noisy meeting audio on UTMOS. We read these two numbers only as evidence that the audio is free of TTS artifacts, so \textbf{model failures cannot be attributed to audio quality}. Speaker embeddings also show large, consistent separation (inter-speaker 0.159 vs.\ intra 0.571; separation 0.412, larger than AMI's 0.299), confirming speakers are clearly distinguishable.
    \item \textbf{Controlled simplifications.} Our stimuli differ from real meetings in \textbf{three} deliberate ways: (i)~the complete absence of overlapping speech (0\% vs.\ AMI's 24.5\%); (ii)~the absence of long ``thinking'' pauses (see the gap-duration rows below); and (iii)~the absence of background noise and reverberation (part of why AMI's WER is far higher than ours). Our inter-turn timing for \textit{active} transitions is itself naturalistic.
\end{itemize}

We report gaps $<$ 1~s separately because AMI, as a real meeting corpus, contains a large fraction (56.6\%) of long silences (participants pausing to think) that inflate its overall mean gap to 1.8~s. Restricting to gaps $<$ 1~s isolates comparable active-turn transitions: there MP-Bench (0.40~s) and AMI (0.43~s) are closely matched, indicating that the overall difference reflects our deliberate omission of long pauses rather than unnaturally fast turn-taking.

\subsection{List of Topics for \Scone~Scenario}
\label{subsec:app_topics}
Topic List: Capital Cities, National Currencies, U.S. State Capitals, Planets of the Solar System, Animal Group Names, Animal Baby Names, Human Anatomy Basics, Simple Arithmetic, Colors of the Rainbow, Days of the Week, Months of the Year, Zodiac Signs, Birthstones, National Flags, Famous Landmarks, Continents, Oceans of the World, Major Rivers, Mountain Ranges, US Presidents, Inventions and Inventors, Famous Explorers, World War II Dates, Greek Mythology Gods, Geometric Shapes, Units of Measurement, Freezing and Boiling Points, States of Matter, Primary and Secondary Colors, Musical Instruments, Famous Classical Composers, Popular Nursery Rhymes, Fairy Tales, Disney Movies, Superheroes and Powers, Harry Potter Houses, Star Wars Characters, Game of Thrones Houses, Oscar-Winning Movies, Sitcom Characters, Famous Movie Quotes, Celebrity Real Names, Pop Music Artists, The Beatles Members, Musical Bands and Lead Singers, Sports Team Mascots, Olympic Host Cities, Rules of Soccer, Rules of Basketball, Rules of Baseball, Tennis Scoring, Golf Terms, Bowling Terms, Chess Pieces, Video Game Characters, Pokemon Types, Ingredients in Common Cocktails, Coffee Types, Tea Varieties, Fruit and Vegetable Identification, Spices and Herbs, Types of Pasta, Breakfast Foods, Fast Food Chains, Candy Bars, Soda Brands, Car Brands and Logos, Compass Directions, Time Zones, Social Media Platforms, Internet Slang, Common Idioms, Synonyms and Antonyms, Collective Nouns, Common Phobias, Holidays and Dates, Seasons, Weather Phenomena, Natural Disasters, Types of Clouds, Tree Types, Dog Breeds, Cat Breeds, Farm Animal

\subsection{Tasks Examples}
\label{subsec:app_task_example}

\subsubsection{Understanding Task for \Scone~Scenario}
\begin{llmprompttemplate}{}
Speakers:
Adam (male) - voice: relatively low male pitch, male, slow, relaxed
David (male) - voice: relatively high male pitch, male, nosy
Rose (female) - voice: female, clear, clean

Transcripts:
Adam:Does anyone know the actual difference between cilantro and coriander?
Rose:I am pretty sure they are entirely different species, Adam. Like, one is an herb and one is a spice seed.
David:No, I cook with them all the time. They are the same plant. Cilantro is the leaves and coriander is the seeds.
Rose:That sounds too simple. I think they are cousins, not the same plant.
David:I am telling you, Rose, it is the Coriandrum sativum plant.
Adam:Okay, David is sounding very confident with the Latin names there, but I am still confused.
Rose:Hey AI Assistant. Please answer the question. What is the name of the speaker with a relatively low male pitch?
\end{llmprompttemplate}

\subsubsection{Behavior Task for \Scone~Scenario}
\begin{llmprompttemplate}{}
Speakers:
Adam (male) - voice: relatively low male pitch, male, slow, relaxed
David (male) - voice: relatively high male pitch, male, nosy
Rose (female) - voice: female, clear, clean

Transcripts:
Adam:Does anyone know the actual difference between cilantro and coriander?
Rose:I am pretty sure they are entirely different species, Adam. Like, one is an herb and one is a spice seed.
David:No, I cook with them all the time. They are the same plant. Cilantro is the leaves and coriander is the seeds.
Rose:That sounds too simple. I think they are cousins, not the same plant.
David:I am telling you, Rose, it is the Coriandrum sativum plant.
Adam:Okay, David is sounding very confident with the Latin names there, but I am still confused.
Rose:Let's settle this. AI Assistant, what is the answer and who has the right information here?
\end{llmprompttemplate}

\subsubsection{Behavior Task for Turn-Based Game}
\begin{llmprompttemplate}{Voice Agent should take turn}
Speakers
Leo (male) - voice: relatively low pitch, male, resonant throaty
David (male) - voice: relatively high pitch, male, nosy
Emma (female) - voice: low pitch, female, intonation-rich

Transcripts:
Leo:Hey everyone, I've got a fun mental exercise for us to try. It's a number response game involving two teams.
Emma:Oh? How does it work, Leo?
Leo:It's simple. We have Group 1 and Group 2. Each has a Leader and a Follower. If the Leader of Group 1 says a number, the Follower of Group 2 has to say the next integer.
David:And if the Leader of Group 2 speaks?
Leo:Then the Follower of Group 1 has to respond with the next integer. It's criss-crossed.
David:Okay, sounds easy enough if we pay attention.
Leo:Great. So, I will take the role of Group 1 Leader, and the AI Assistant will be the Group 1 Follower. Emma, you are the Group 2 Leader.
Emma:Understood. That means David is my Follower in Group 2?
Leo:Exactly, David is Group 2 Follower. Everyone ready? Game start!
Leo:5
David:6
Leo:12
\end{llmprompttemplate}

\begin{llmprompttemplate}{Voice Agent shouldn't take turn}
Speakers
Leo (male) - voice: relatively low pitch, male, resonant throaty
David (male) - voice: relatively high pitch, male, nosy
Emma (female) - voice: low pitch, female, intonation-rich

Transcripts:
Leo:Hey everyone, I've got a fun mental exercise for us to try. It's a number response game involving two teams.
Emma:Oh? How does it work, Leo?
Leo:It's simple. We have Group 1 and Group 2. Each has a Leader and a Follower. If the Leader of Group 1 says a number, the Follower of Group 2 has to say the next integer.
David:And if the Leader of Group 2 speaks?
Leo:Then the Follower of Group 1 has to respond with the next integer. It's criss-crossed.
David:Okay, sounds easy enough if we pay attention.
Leo:Great. So, I will take the role of Group 1 Leader, and the AI Assistant will be the Group 1 Follower. Emma, you are the Group 2 Leader.
Emma:Understood. That means David is my Follower in Group 2?
Leo:Exactly, David is Group 2 Follower. Everyone ready? Game start!
Leo:5
David:6
Emma:12
\end{llmprompttemplate}

\begin{table*}[t]
\small
\begin{tabular}{lrrrr}
\toprule
\textbf{Setting} & \textbf{Und. Overall Acc.} & \textbf{Spk. Name Acc.} & \textbf{Disc. App Acc.} & \textbf{TB TT Acc.} \\
\midrule
Distinct voices & 40.0\% (20/50) {\tiny[27.6, 53.8]} & 45.0\% (18/40) {\tiny[30.7, 60.2]} & 75.0\% (15/20) {\tiny[53.1, 88.8]} & 50.0\% (15/30) {\tiny[33.2, 66.8]} \\
Similar voices  & 4.0\% (2/50) {\tiny[1.1, 13.5]}   & 0.0\% (0/40) {\tiny[0.0, 8.8]}    & 45.0\% (9/20) {\tiny[25.8, 65.8]}  & 43.3\% (13/30) {\tiny[27.4, 60.8]}  \\
\bottomrule
\end{tabular}
\caption{Covo-Audio, distinct- vs.\ similar-voice subsets (95\% Wilson intervals in \%). Disc.\ App = Discussion appropriateness; TB TT = Turn-based turn-taking.}
\label{tab:similar_voice_ablation}
\end{table*}

\section{Details of Human Evaluation}
\label{sec:app_human_eval}
We recruit 13 graduate-student annotators from two universities with backgrounds in speech and language processing.
All annotators agree to participate in this human evaluation study and they agree their annotation data will be used for this study.
Each annotator works on a batch of 30 items per session.
The annotation guideline instructs that a response is considered consistent only if it (1)~explicitly states the model's own claim or answer and (2)~explicitly identifies the speaker whose claim it agrees with; if either element is missing the response is judged inconsistent.

\section{Robustness of LLM Judge across model families}
\label{sec:app_robustness_llmjudges}
To test whether our GPT-5 labels are idiosyncratic, we re-judged a stratified subsample of \textbf{720} items (Understanding--Discussion: 14 settings $\times$ 30; Behavior--Discussion: 6 settings $\times$ 50; seed 20260712) with three additional judges from different families. (Other metrics like Turn-Taking is judged by timestamps so is independent of the LLM judge.)

\begin{table}[h]
\centering
\small
\begin{tabular}{lrrr}
\toprule
\textbf{Judge} & \textbf{$n$} & \textbf{Exact match} & \textbf{Cohen's $\kappa$} \\
\midrule
Gemini-3.1-Flash-Lite & 720 & 88.9\% & 0.78 \\
Gemini-3.1-Pro        & 720 & 88.8\% & 0.77 \\
Claude Opus 4.6       & 720 & 91.7\% & 0.83 \\
\bottomrule
\end{tabular}
\caption{Agreement across LLM judges.}
\label{tab:judge_agreement}
\end{table}

By scenario, Understanding is especially stable ($\kappa$ = 0.83 / 0.85 / 0.90 for Flash-Lite / Pro / Claude) and Behavior remains substantial ($\kappa$ = 0.74 / 0.71 / 0.78). All three judges are only mildly more lenient than GPT-5, with no systematic flips that would reorder model rankings.

\section{Ablation for high voice similarity speakers for MP-Bench}
\label{sec:app_similar_voices}

In our benchmark, we chose distinct voices deliberately, for two reasons:

\begin{enumerate}
    \item As a first step, MP-Bench targets the \textit{core} multiparty ability, so we disentangle it from speaker-diarization difficulty; distinct voices keep diarization from being the bottleneck.
    \item It enables \textbf{objective, unambiguous} question design (e.g., ``\textit{what did the higher-pitched speaker suggest?}''), which is hard to grade objectively when voices are too similar.
\end{enumerate}

As a ablation, we additionally ran a \textbf{similar-voice ablation} on the strongest realtime understanding model (Covo-Audio), excluding questions that explicitly reference pitch to avoid ambiguity.

As shown in Table~\ref{tab:similar_voice_ablation}, when voice separability is reduced, the abilities that depend on acoustic attribution collapse: Understanding drops 40\% to 4\%, speaker-name accuracy drops 45\% to 0\%, and Discussion appropriateness drops 75\%$\rightarrow$45\%. Turn-based turn-taking remains near/below the 50\% chance baseline throughout. Thus our negative findings of weak multiparty understanding and near-chance implicit/negative turn-taking \textbf{persist, and for understanding become more severe,} as voices become harder to tell apart.

\section{LLM Prompt Templates}
\label{sec:app_llm_prompt}
\subsection{Inference-Time System Prompts for Realtime Models}
\label{sec:app_realtime_prompts}

The realtime audio models in our experiments are driven by one of the two short voice-assistant system prompts shown below.
The \emph{default} prompt is used in all main results~(Table~\ref{tab:main_bench_results}); the \emph{multi-party-aware} variant---which differs only in a single added sentence~(``Be aware that there may be multiple speakers in the conversation.'')---is used solely for the prompt ablation in Table~\ref{tab:ablation_prompt}.

\begin{llmprompttemplate}{Default System Prompt}
You are a helpful voice AI assistant. Keep your responses concise and conversational since they will be spoken aloud. Respond naturally to whatever the user says or asks.
\end{llmprompttemplate}

\vspace{1em}

\begin{llmprompttemplate}{Multi-Party-Aware System Prompt}
You are a helpful voice AI assistant. Be aware that there may be multiple speakers in the conversation. Keep your responses concise and conversational since they will be spoken aloud. Respond naturally to whatever the users say or ask.
\end{llmprompttemplate}

\vspace{1em}

\subsection{LLM Prompt for Scenario Generation}

\begin{llmprompttemplate}{\Scone~Scenario Generation}
You are a helpful assistant.
Please generate {num_diags_per_generation} naturally spoken 3 people conversation.
The names of the 3 people are:
{speaker_names_str}

In the beginning of the conversation, the first person ask a factual question based on the topic of "{topic_list_str}".
Then, all people give their answers in random order. (Speaker turns interleaved each other)
At the end, one person redirect the question to "AI Assistant" and ask it to judge who is correct.
The name of the AI assistant should be called "AI Assistant". Please regard "AI Assistant" as a human.
Make sure that all speaker names are only mentioned once in one of the speaker's spoken content during the conversation. 
The same speaker can speaks multiple turns.

The output is a list of generated conversations.
Each conversation contains:
1. Topic of the factual question asked by the first person.
2. conversation (a list of speaker names together with utterances)

The output should follow the JSON format.
[{{"topic": ...,  conversation": [{{"speaker": ..., "content": ... }}, {{"speaker": ..., "content": ... }}, ...]}}, {{"topic": ...,  conversation": [{{"speaker": ..., "content": ... }}, {{"speaker": ..., "content": ... }}, ...]}} ...]
\end{llmprompttemplate}

\vspace{1em}

\begin{llmprompttemplate}{\Sctwo~Scenario  Generation}
You are a helpful assistant.
Please generate {num_diags_per_generation} naturally spoken 4 people conversation.
The speakers are: {speaker_names_str} and "AI Assistant". Please regard "AI Assistant" as a human participant too.

In the conversation, people are playing a game.
In the game, two people formed a group so we have Group1 and Group2.
In each group, one person is the leader and the other is the follower.
{game_rules}

In the beginning of the conversation, the first person(should not be "AI Assistant") initiate to play the game and also introduce the rules.
Afterwards, the first person assign people into Group1, Group2 and also leaders and followers.
During the introduction of game rule, the two other people (excluding "AI Assistant") could also interact with the first person.
After that, the first person anounces game start without reveal anything related to who start the game first.
Throughout the generated conversation, "AI Assistant" should not speak.

Given the following group and role assignment:
Group1 Leader: {group1_leader}
Group1 Follower: {group1_follower}
Group2 Leader: {group2_leader}
Group2 Follower: {group2_follower}


Please generate the conversation until the first person anounces game start.
{game_pairs_instruction}
Make sure that all speakers (excluding "AI Assistant")'s name are being addressed in the game rule introduction phase.
The same speaker can speaks multiple turns.


The output contains a list of generated spoken conversations and each example contains the list of speaker names, answer pairs and a list of utterances of the conversation.
The output should follow the JSON format.
[{{"speaker_names:" ["...", "...", "..."], "answer_pairs": [["...", "..."], ...] "conversation": [{{"speaker": ..., "content"... }}, {{"speaker": ..., "content"... }}, ...]}}, {{"speaker_names:" ["...", "...", "..."], "answer_pairs": [["...", "..."], ...] "conversation": [{{"speaker": ..., "content"... }}, {{"speaker": ..., "content"... }}, ...]}},...]
\end{llmprompttemplate}

\subsection{LLM Prompt for Validating Generated Main Dialogues}
\begin{llmprompttemplate}{Validation without Speaker Names}
You are a helpful assistant.
You are given a multiparty spoken conversation transcript.
Please predicted the speaker names of each utterances.

Input format (JSON):
[{{"content": ...}}, {{"content": ...}}, ...]

Output format (JSON):
[{{"speaker": ..., "content": ...}}, {{"speaker": ..., "content": ...}}, ...]

Now given:
Input:
{conversation_transcript_wo_spks}

Output:
\end{llmprompttemplate}
\vspace{1em}
\begin{llmprompttemplate}{Validation with Anonymized Speaker Names}
You are a helpful assistant.
You are given a multiparty spoken conversation transcript.
For each utterances, the speaker's names are anonymized as Speaker1, Speaker2, Speaker3.
Please predicted the real speaker names of Speaker1, Speaker2, Speaker3 based on the given information.

Input format (JSON):
[{{"speaker_anonymized": "...", content": ...}}, {{"speaker_anonymized": "...", "content": ...}}, ...]

Output format (JSON):
{{"Speaker1": "...", "Speaker2": "...", "Speaker3": "..."}}

Now given:
Input:
{conversation_transcript_w_anonymized_spks}

Output:
\end{llmprompttemplate}

\subsection{LLM-as-a-Judge Prompts}
\label{app:llm_judge_prompts}

This subsection lists the four LLM-as-a-judge prompts that produce the numbers reported in the main table and ablations.
All four are run with GPT-5 as the judge.
The first prompt grades the \emph{Understanding} task, the next two grade the two \emph{Behavior} sub-tasks (\emph{\Scone} consistency and \emph{Turn-Based} rule compliance), and the last is the dedicated fuzzy-matching judge used to compute Spk.\ Name Acc.\ on the speaker-name slice of the Understanding QA pool.

\vspace{1em}

\begin{llmprompttemplate}{Understanding QA Judge}
You are a helpful assistant.
You are given a meta-transcript of a multiparty spoken conversation together with a question, ground truth answer and a predicted answer.
The question is related to the spoken conversation.
The meta-transcript includes the speakers (name,gender,voice_characteristics) that are in the conversation.
The meta-transcript also includes each spoken turn in the conversation. Each turn contains information for speaker, content, addressee and coreferences.
Please judge the correctness of the predicted answer.
If the predicted answer says something like there is not enough information or refuse to give an answer, please consider it as incorrect.

The input is given in the following format:
Metatranscripts for Multiparty Spoken Conversation:
{{"speakers": [{{"name": "...", "gender": "...", "voice_characteristics": "..."}}, ...], "metatranscript": [{{"speaker": "...", "content": "...", "addressee": ["...", ...], "coreferences": {{"...": "...", ...}}}}, ...]}}
Question:
...
Ground Truth Answer:
...
Predicted Answer:
...

Please output with the following format:
{{"correct": true/false}}

Now given
Metatranscripts for Multiparty Spoken Conversation:
{metainformation}
Question:
{question_text}
Ground Truth Answer:
{answer_text}
Predicted Answer:
{llm_predicted_answer}

Output:
\end{llmprompttemplate}

\vspace{1em}

\begin{llmprompttemplate}{\Scone~Consistency Judge}
You are a helpful assistant.
You are given a meta-transcript of a multiparty spoken conversation together with a model response
The model's name is "AI Assistant" and is one of the participant in the spoken conversation.
The meta-transcript includes the speakers (name,gender,voice_characteristics) that are in the conversation.
The meta-transcript also includes each spoken turn in the conversation. Each turn contains information for speaker, content, addressee and coreferences.
Please judge whether the model's response is consistent with the spoken conversation.
The model's response is considered as consistent if the person it identified as correct also have the same claim as its own answer.
Do not judge the consistency of the model's response based on factual correctness.

The input is given in the following format:
Metatranscripts for Multiparty Spoken Conversation:
{{"speakers": [{{"name": "...", "gender": "...", "voice_characteristics": "..."}}, ...], "metatranscript": [{{"speaker": "...", "content": "...", "addressee": ["...", ...], "coreferences": {{"...": "...", ...}}}}, ...]}}
Model's response:
...

Please output with the following format:
{{"consistent": true/false}}

Now given
Metatranscripts for Multiparty Spoken Conversation:
{metatranscript}
Model's response:
{llm_predicted_answer}

Output:
\end{llmprompttemplate}

\vspace{1em}

\begin{llmprompttemplate}{Turn-Based Game Judge (with chain-of-thought reasoning)}
You are a helpful assistant.
You are given a transcript of a multiparty spoken conversation together with a model response.
The model's name is "AI Assistant" and is one of the participants in the spoken conversation.
The people in the conversation are playing a turn-based game. The game rules are described within the conversation itself.
The transcript includes the speakers (name, gender, voice_characteristics) and each spoken turn (speaker, content).

Your task is to judge whether the model's response is consistent with the game being played.
You MUST follow these steps in order and include your reasoning:

Step 1: Identify the game and its rules from the conversation.
Step 2: Identify ALL group/team assignments -- specifically, who is the Leader and who is the Follower of each group.
Step 3: Determine which Leader's speech triggers the AI Assistant to respond. In a cross-group reaction game, Group X's Leader triggers Group Y's Follower (NOT Group X's own Follower).
Step 4: Look at the LAST speaker before the model's response. Determine whether that speaker is the Leader who triggers the AI Assistant.
Step 5: Based on Step 4:
  - If the last speaker IS the Leader who triggers the AI Assistant -> the AI SHOULD respond. Check if the response follows the game rules (e.g., says a random word in a word game).
  - If the last speaker is NOT the Leader who triggers the AI Assistant -> the AI should remain SILENT. Any non-empty response means it acted out of turn.
  - An empty response means the model remained silent.

Focus ONLY on the model's final response content. Even if the model's reasoning or explanation contains errors, as long as the final action follows the game rules, consider it consistent.
Do not judge the model's response based on factual correctness outside of the game context.

=== EXAMPLE 1 (consistent = true) ===

Transcript:
{{"speakers": [{{"name": "Alice", "gender": "female", "voice_characteristics": "clear"}}, {{"name": "Bob", "gender": "male", "voice_characteristics": "deep"}}], "metatranscript": [{{"speaker": "Alice", "content": "OK new plan. Two groups. Group 1: Leader = Alice, Follower = Bob. Group 2: Leader = Bob, Follower = AI Assistant. When Group 1's Leader speaks a word, Group 2's Follower must respond with a random word. When Group 2's Leader speaks, Group 1's Follower responds. Ready? Go!"}}, {{"speaker": "Alice", "content": "Sun"}}]}}
Model's response: Moon

Reasoning:
- Game: Cross-group word reaction game.
- Group 1: Leader = Alice, Follower = Bob.
- Group 2: Leader = Bob, Follower = AI Assistant.
- Trigger rule: Group 1's Leader (Alice) triggers Group 2's Follower (AI Assistant).
- Last speaker before model's response: Alice (Group 1's Leader).
- Alice IS the Leader who triggers the AI Assistant -> the AI SHOULD respond.
- The AI responded with "Moon", which is a valid random word.
- Verdict: consistent = true.

=== EXAMPLE 2 (consistent = false) ===

Transcript:
{{"speakers": [{{"name": "Lily", "gender": "female", "voice_characteristics": "clear"}}, {{"name": "David", "gender": "male", "voice_characteristics": "warm"}}, {{"name": "Jack", "gender": "male", "voice_characteristics": "deep"}}], "metatranscript": [{{"speaker": "Lily", "content": "Let's play a reaction game! Group 1: Leader = Lily, Follower = David. Group 2: Leader = Jack, Follower = AI Assistant. If Group 1's Leader speaks, Group 2's Follower responds. If Group 2's Leader speaks, Group 1's Follower responds. Go!"}}, {{"speaker": "Jack", "content": "Apple"}}]}}
Model's response: Banana

Reasoning:
- Game: Cross-group word reaction game.
- Group 1: Leader = Lily, Follower = David.
- Group 2: Leader = Jack, Follower = AI Assistant.
- Trigger rule: Group 1's Leader (Lily) triggers Group 2's Follower (AI Assistant). Group 2's Leader (Jack) triggers Group 1's Follower (David).
- Last speaker before model's response: Jack (Group 2's Leader).
- Jack is the AI Assistant's OWN group's Leader. Jack's speech triggers David (Group 1's Follower), NOT the AI Assistant.
- The AI should have remained silent, but it responded with "Banana".
- Verdict: consistent = false.

=== END OF EXAMPLES ===

The input is given in the following format:
Transcript for Multiparty Spoken Conversation:
{{"speakers": [{{"name": "...", "gender": "...", "voice_characteristics": "..."}}, ...], "metatranscript": [{{"speaker": "...", "content": "..."}}, ...]}}
Model's response:
...

Please output a JSON object with your reasoning and judgment:
{{"reasoning": "your step-by-step analysis here", "consistent": true/false}}

Now given
Transcript for Multiparty Spoken Conversation:
{metatranscript}
Model's response:
{llm_predicted_answer}

Output:
\end{llmprompttemplate}

\vspace{1em}

\begin{llmprompttemplate}{Speaker-Name Fuzzy Match Judge (system prompt)}
You are a strict evaluator for a speaker-name identification task.

The model was asked a question about a multiparty spoken conversation (e.g., "What is the name of the speaker with a warm voice?"). The ground truth answer is a single speaker name.

The model heard the conversation as audio and responded via speech, which was then transcribed by ASR. The ASR transcription may contain errors: misspelled names (e.g., "Karen" -> "Caring", "Cairn"; "Leo" -> "Lioh"; "Lily" -> "Leely"), foreign-language greetings, filler words, or garbled text.

Your task:
1. Read the model's transcribed response carefully.
2. Determine what speaker name the model intended as its answer.
   - Ignore greetings, hedging, filler, or refusal disclaimers.
   - If the model mentions ANY valid speaker name anywhere in the response -- even if surrounded by refusal language like "I can't identify speakers" -- extract that name.
   - If the model mentions multiple names, take the one most likely intended as the answer to the question (often the last or most confidently stated).
   - Account for phonetic / ASR spelling errors (e.g., "Caring" = "Karen", "Lioh" = "Leo").
3. Compare the extracted name to the ground truth.

IMPORTANT: A response like "I can't identify speakers, but it sounds like it might be Karen" should extract "Karen" -- the refusal disclaimer does NOT prevent name extraction. Only set extracted_name to null if absolutely NO valid speaker name (or phonetic variant) appears anywhere in the response.

Valid speaker names: Adam, Anna, Ben, David, Emma, Jack, Karen, Leo, Lily, Luke, Mia, Rose.

Output EXACTLY one JSON object (no markdown):
{{"extracted_name": "<name or null>", "correct": true/false, "reasoning": "<brief>"}}

Rules:
- "extracted_name" must be one of the 12 valid names above, or null if the model gave no identifiable name guess anywhere in the response.
- "correct" is true only when extracted_name matches the ground truth (case-insensitive).
- Only set extracted_name to null if NO valid name or phonetic variant appears in the response at all.
\end{llmprompttemplate}

\vspace{1em}

\begin{llmprompttemplate}{Speaker-Name Fuzzy Match Judge (user message)}
Question: {question}
Ground truth answer: {gt_name}

Model's transcribed response (may contain ASR errors):
{response}
\end{llmprompttemplate}

\section{Use of Generative AI Tools}
\label{sec:app_ai_usage}

We used generative AI assistants in three supporting roles: polishing the
writing, assisting with implementation and debugging
of our evaluation and analysis code, and assisting with figure and table
preparation. 
AI tools were not used to originate the research idea, design
the benchmark, or interpret the results. 
All AI-assisted text, code, and
figures were reviewed by the authors, who take full responsibility for the
motivation, methodology, results, and any errors in this paper.

\end{document}